# Akhiezer Mechanism Dominates Relaxation of Propagons in Amorphous at Room Temperature

Yuxuan Liao[1], and Junichiro Shiomi[1] [*]

[1]*Department of Mechanical Engineering, The University of Tokyo, 7-3-1 Hongo, Bunkyo, Tokyo 113-8656, Japan*

[*]E-mail: shiomi@photon.t.u-tokyo.ac.jp

Propagons play an important role in tuning the thermal conductivity of nanostructured amorphous materials. Although advances have been made to quantitatively evaluate the relaxation time of propagons with molecular dynamics, the underlying relaxation mechanism remains unexplored. Here, we investigate the relaxation process of propagons in amorphous silicon, amorphous silica, and amorphous silicon nitride at room temperature in terms of Akhiezer model, the parameters of which were evaluated by performing lattice dynamics and molecular dynamics analysis. The results show that the Akhiezer model can well reproduce experimental results obtained by various kinds of measurement methods, indicating that Akhiezer mechanism dominates the relaxation process of propagons at room temperature. Moreover, we show that the appropriate sound speed of propagons is around 80% of the Debye sound speed and comparable to that of the sound speed of the transversal modes. We also reveal that the contribution of diffusons to the total thermal conductivity of these amorphous is similar, which is around 1 W/m K, while the contribution of propagons varies significantly depending on the materials, which is 30% in amorphous silicon and silica but can be as high as 70% in amorphous silicon nitride.

I. INTRODUCTION

The thermal conductivity ($\kappa$) of semiconductors and insulators, which is governed by vibrational mode of atoms, has been one of the key parameters that determine the usability and functionality of the materials in a wide range of applications. In crystal materials, these vibrational modes can be quantized as quasiparticles referred to as phonons. One of the most effective ways to manipulate the $\kappa$ of the crystal is to scatter thermal phonons by introducing nanostructures with length scales comparable to their intrinsic mean free paths (MFPs), which usually vary from several nanometers to tens of nanometers at room temperature, depending on the material. The experimentally achieved $\kappa$ of a nanostructured material through nanotechnology can be smaller by orders of magnitude than that of the bulk [1-5]. By assuming phonons as particles, the measured $\kappa$ of various kinds of bulk and nanostructured crystal materials can be well reproduced using the Boltzmann transport equation (BTE) [6-9] with recent advances in first-principles calculations [5,10-13].

By contrast, manipulating the $\kappa$ of amorphous material through nanostructuring is much more challenging. This is because the $\kappa$ of amorphous was thought to be dominated by diffusons, which exhibit random walks in the material with step lengths comparable to atomistic spacings and thus can hardly be scattered by nanostructures, as indicated in the theory of minimum thermal conductivity and Allen and Feldman (AF) [14-16]. Despite the fact that phonon-like vibrational modes (propagons) with low-frequencies and long effective MFPs exist in amorphous, their contribution to $\kappa$ was neglected because of their very limited frequency range and density of states [16].

However, recent theoretical works of Larkin *et al* [17] and Zhou *et al* [18] have shown that propagons in amorphous silicon (a-Si) can have long MFPs that range from tens of nanometers to several micrometers, and thus propagons could contribute to about 40% of the total $\kappa$ of the bulk [17,18]. Experimental works favor their theoretical analysis, in which 50% of $\kappa$ reduction has been reported in the cross-plane direction of a-Si thin films [19] and in-plane directions of suspended nanostructures [20-22]. These findings strongly suggest that propagons in fact are able to contribute to a relatively large part of the total $\kappa$ of amorphous, and thus should be considered when analyzing the heat conduction, despite their small density of states.

The fact that propagons could contribute to a large part of the total $\kappa$ offers the possibility to further reduce already-small $\kappa$ of amorphous by introducing nanostructures to scatter propagons with long MFPs in a way similar to that in crystal materials, which is of crucial importance for industrial applications such as thermal barriers and insulators. Despite the importance of the knowledge in the MFP of propagons, the mechanism of attenuation or relaxation that determines the MFP remains to be explored. Investigations on the MFP of propagons up to date only rely on the computationally intensive normal-mode-decomposition-based molecular dynamics (NMD) [17], which can hardly reveal the underlying relaxation mechanisms. It therefore calls for a theoretical model that not only gives the magnitude of the relaxation time, but also identifies the relaxation process of propagons.

Here, by using a-Si, amorphous silica (a-SiO$_2$), and amorphous silicon nitride (a-Si$_3$N$_4$) as examples, we demonstrated that the room-temperature relaxation process of propagons is dominated by Akhiezer mechanism, which is a coupling of the strain of sound waves and thermal vibration modes [23-25]. The relaxation time of propagons predicted by the Akhiezer model is able to well reproduce experimental results obtained by various kinds of measurement methods. The contribution of propagons to the $\kappa$ of amorphous in terms of relaxation time obtained from the Akhiezer model is also discussed in detail.

The sections of the paper are organized as follows. The theoretical frameworks in terms of propagons and diffusons transport are discussed in Sec. II. The numerical sample preparation of these amorphous is discussed in Sec. III. The calculation of relaxation time in terms of the Akhiezer model is shown in Sec. IV. A discussion of the $\kappa$ of amorphous is summarized in Sec. V, where a comparison was made with previous works and experimental measurements.

## II. THEORY FOR THERMAL CONDUCTIVITY OF AMORPHOUS MATERIALS

The total thermal conductivity ($\kappa_\text{T}$) of the amorphous materials includes the contribution of both propagons ($\kappa_\text{P}$) and diffusons ($\kappa_\text{D}$):

$$\kappa_\text{T} = \kappa_\text{P} + \kappa_\text{D} \tag{1}$$

where $\kappa_\text{P}$ from phonon-like propagons follows the phonon gas model [17]:

$$\kappa_\text{P} = \frac{1}{3}\int_0^{\omega_\text{t}} C(\omega)v_s^2\tau(\omega)d\omega \tag{2}$$

and $\kappa_\text{D}$ from diffusons is evaluated by the AF theory [15]:

$$\kappa_\text{D} = \int_{\omega_\text{t}}^{\infty} C(\omega)D(\omega)d\omega \tag{3}$$

where $\omega$ is frequency, and $\omega_\text{t}$ is the transition frequency of propagons and diffusons. $C(\omega)$ is the volumic specific heat capacity absorbs the vibrational-mode density of states, $v_s$ is the appropriate sound speed of propagons, $\tau(\omega)$ is the vibrational-mode relaxation time, and $D(\omega)$ is the frequency-dependent diffuson diffusivity from AF theory.

The $D(\omega)$ in mode-dependent fashion is expressed as:

$$D(\omega_i) = \frac{\pi V^2}{3\hbar^2 \omega_i^2} \sum_{j \neq i} (|S_{ij}|)^2 \delta(\omega_i - \omega_j) = \frac{\pi}{3} \sum_{j \neq i} \left(\frac{v_{ij}(\omega_i + \omega_j)}{2\omega_i}\right)^2 \delta(\omega_i - \omega_j) \tag{4}$$

where $i$ and $j$ are the index of vibrational modes; $\hbar$ is the reduced Planck constant; $V$ is the volume of the system; $\delta$ is the delta function broadened into Lorentzian with a broadening width as $5\delta\omega_{avg}$ ($\delta\omega_{avg}$ is the average mode frequency spacing); $S_{ij}$ is the $(i,j)^{th}$ element of the heat current operator; $v_{ij}$ is the $(i,j)^{th}$ element of the Hardy's group velocity operator [26].

The Hardy's group velocity operator can be applied to both crystal and amorphous, and it is expressed as [26]:

$$v_{ij} = \frac{i}{2\sqrt{\omega_i \omega_j}} \sum_{\alpha,\beta,m,l} e_{\alpha,m,i} H_{\alpha\beta}^{ml} R_{ml} e_{\beta,l,j} \tag{5}$$

where $\alpha, \beta$ are the index of orientations; $m$ and $l$ label the atoms; $R_{ml}$ is the distance between atom $m$ and $l$; $H$ and $e$ are the mass-scaled Hermitian force constant and eigenvectors, respectively.

In crystals, $v_{ii}$ represents phonon group velocity, and $v_{ij}$ ($i \neq j$) vanishes because the $i^{th}$ and $j^{th}$ eigenvectors are orthogonal. While in amorphous, vibrational modes are defined only at Gamma points, thus, the $v_{ii}$ terms are intrinsically 0, leaving only $v_{ij}$ ($i \neq j$) terms dominate thermal transport, as expressed by Eq. (4).

The element $v_{ij}$ in amorphous clearly illustrates the character of the heat transfer between the $i^{th}$ and $j^{th}$ modes: the energy of the excited $i^{th}$ mode is projected onto the direction of all the $j^{th}$ modes. Since the eigenvectors of diffusons are randomized, the energy of the $i^{th}$ mode diffusively dissipates. By contrast, the eigenvectors of propagons are aligned [27], thus, the heat carried by propagons propagates in a similar manner as that for phonons.

The calculation of the mode-dependent diffusivity in bulk amorphous is well established, as shown in the works of Allen *et al*. [15] and Larkin *et al* [17]. Here, we focus on propagons, the effective dispersion of which is linear, and thus their density of states (*DOS*) is approximated by the Debye approximation:

$$DOS(\omega) = \frac{3V\omega^2}{2\pi^2 v_s^3} \tag{6}$$

The relaxation time of propagons can be obtained by NMD calculations, or by using the Akhiezer model:

$$\tau_A^{-1} = \frac{C_v T \tau_{av}(\langle\gamma^2\rangle - \langle\gamma\rangle^2)}{\rho v_s^2} \cdot \omega^2 = B\omega^2 \tag{7}$$

with:

$$\langle\gamma^2\rangle - \langle\gamma\rangle^2 = \sum_i \frac{C_i \gamma_i^2}{\sum_i C_i} - \sum_i \left(\frac{C_i \gamma_i}{\sum_i C_i}\right)^2 = \gamma_{av} \tag{8}$$

where $C_v$ is specific heat; $T$ is 300 K as the room temperature; $B$ is the magnitude of relaxation time that defined by Eq (7); $\tau_{av}$ is the average relaxation time of vibration modes; $C_i$ and $\gamma_i$ are the mode-dependent heat capacity and Grüneisen parameter, respectively.

## III. PREPARATION OF AMORPHOUS MATERIALS

Firstly, we constructed the required bulk amorphous structures by the melt-quenching method using molecular dynamics (MD) with the Tersoff interatomic potentials [28,29] in the LAMMPS package [30]. The parameters of Tersoff potential for a-Si, a-SiO$_2$ and a-Si$_3$N$_4$ are respectively taken from the Si(C) set in TABLE I in the works of Tersoff *et al* [28], Table 1 in the works of S. Munetoh *et al* [31], and Table 1 in the works of F. de Brito Mota *et al* [32]. We choose these parameters because they agree well with experimental measurements and the results obtained from first-principles calculations. In these first-principles calculations, the local density approximation (LDA) for the exchange and correlation functional is used for the a-Si [28] system and a-Si$_3$N$_4$ system[32,33], whereas generalized gradient approximation (GGA) for the exchange-correlation term is used for the a-SiO$_2$ system [31].

As for the melt-quenching method for a-Si, a 4096-atom cubic silicon crystal with an equal side length of 4.344 nm was melted at 3500 K. The liquid silicon was then quenched from 3500 K to 300 K with a quenching rate of 0.05 K/ps in an *NPT* ensemble. The structure was then relaxed at 300 K in an *NPT* ensemble for 10 ns and *NVT* ensemble for 10 ns to reduce the residual stress and strain. Finally, energy minimization was performed to obtain stable a-Si structures. With the same procedure, we prepared the a-SiO$_2$ structures, which is an 8748-atom cubic structure with a side length of 5.04 nm, and the a-Si$_3$N$_4$ structure, which is a 7560-atom cuboid structure with side lengths of 4.62 nm, 3.98 nm, and 4.41 nm.

The timesteps in the MD calculations for a-Si, a-SiO$_2$, and a-Si$_3$N$_4$ are respectively set to 0.5 fs, 0.2 fs, and 0.1 fs to resolve their maximum vibration frequency. Note that during the quenching process for a-Si$_3$N$_4$, strong N-N bonds inevitably form, which dramatically decreases the stability of the a-Si$_3$N$_4$ structures. Therefore, we turn off the attractive forces for Si-Si and N-N bond in the quenching procedure such that only Si-N chemical bonds can form until the temperature goes down to 1000 K. After that, we turn on the attractive forces for the cooling, relaxation, and energy minimization procedures. This artificial setup helps us obtain high-quality a-Si$_3$N$_4$ structures.

The obtained atomistic structures of these amorphous are shown in Fig. 1, the corresponding radial distribution functions of them clearly show that they are indeed amorphous materials. By identifying the first and second peaks of the radial distribution functions, we checked that the calculation shows good agreement with the experimental data [34-36] and the previous calculation result [17,36]. Moreover, the densities of the obtained a-Si$_3$N$_4$, a-SiO$_2$, and a-Si are 2800 kg/m$^3$, 2230 kg/m$^3$, and 2300 kg/m$^3$ respectively, which match well with theoretical and measured results [17,37].

## IV. MODAL VIBRATION PROPERTIES OF BULK AMORPHOUS MATERIAL
### IV(A). Vibrational density of states and appropriate sound velocity

After the preparation of these amorphous structures, we are able to obtain the eigenmodes and frequencies of vibrational modes at Gamma points by lattice dynamics using GULP [38]. The frequency-dependent density of states (*DOS*) is then computed from:

$$DOS(\omega) = \sum_i \delta(\omega_i - \omega) \tag{9}$$

where the delta function is approximated by the step function with the energy width as 3 meV.

Fig. 2 plots the obtained *DOS* for a-Si, a-SiO$_2$, and a-Si$_3$N$_4$, which clearly shows the $\omega^2$-dependent trend at low frequencies, indicating that the Debye approximation is applicable. Based on the $\omega^2$-dependent trend in *DOS*, $\omega_t$ for propagons is determined as 4 THz for a-Si$_3$N$_4$, 1.3 THz for a-SiO$_2$, and 2 THz for a-Si.

By fitting the data of *DOS* for propagons with Eq. (6), we obtained the appropriate sound velocities ($v_s$) for a-Si, a-SiO$_2$, and a-Si$_3$N$_4$ (Table 1). The value of $v_s$ for a-Si and a-SiO$_2$ matches well with those reported by Larkin *et al* [17], within an error less than 12%. Since originally, the velocity term in the Debye *DOS* is the Debye sound speed ($v_{dby}$), it is natural to compare $v_s$ with $v_{dby}$. The Debye sound speed is estimated from the sound speed of longitudinal ($S_L$) mode and transversal mode ($S_T$) as:

$$\frac{3}{v_{dby}^3} = \frac{1}{S_L^3} + \frac{2}{v_T^3} \tag{10}$$

In order to obtain $v_{dby}$, we calculated $S_L$ and $S_T$ for the three materials, which is further compared with experimental measurements and previous works (Table 1). The calculated $S_L$ (=7915 m/s) and $S_T$ (=4198 m/s) for a-Si in our work agree with the values obtained by molecular dynamics calculations of Larkin *et al* [17] and by experimental measurements using Brillouin scattering [39]. For a-SiO$_2$, our prediction ($S_L$: 6200 m/s, $S_T$: 3234 m/s) is larger than those in the works of Larkin *et al* [17], however, the error (23% for $S_L$, 15% for $S_T$) is acceptable when considering the fact that Larkin *et al* have used the Beest-Kramer–van Santen (BKS) potential. On the other hand, our results agree better with experimental data measured by inelastic x-ray scattering [40-43], suggesting that the Tersoff potential works better for predicting the sound speed of a-SiO$_2$ than BKS potential. The $S_L$ (=12000 m/s) and $S_T$ (=6680 m/s) for a-Si$_3$N$_4$ also agree well with the experimental values obtained by using ultrasonic measurements [37], with the error as 14% for $S_L$ and 7% for $S_T$.

Following the validation of the calculation of $S_L$ and $S_T$, the $v_{dby}$ is obtained by Eq. (10). The results are summarized in Table 1. It shows that $v_s$ is smaller than $v_{dby}$, with an error varying from 13.3% to 22.6% (18.6% on average). The reason is that viscosity damping lowers $v_s$ of propagons, as revealed in the works of G. Baldi *et al* [44], Rat *et al* [45] and Vacher *et al* [46]. Additionally, we noticed that $v_s$ is comparable to $S_T$ for these amorphous materials (Table 1), with an error smaller than 13%. Thus, we can use $S_T$ to approximate $v_s$ for amorphous materials.

**IV(B). Vibrational mode Grüneisen parameter of amorphous**

The vibrational mode-dependent Grüneisen parameter is calculated by its definition as:

$$\gamma = \frac{d\omega/\omega}{dV/V} \tag{11}$$

where $dV/V$ is the definition of strain, and $d\omega/\omega$ is the corresponding perturbation on energy or frequency of each vibration mode.

Numerically, the side length of the prepared bulk amorphous samples is expanded by 1%, which gives the relative change in volume $dV/V$ as $(1+0.01)^3-1=3.03\%$. The small change of the volume gives perturbation on the energy of each vibration mode by $d\omega/\omega$, which can be obtained by lattice dynamics. The calculated mode-dependent $\gamma$ for a-Si, a-SiO$_2$ and a-Si$_3$N$_4$ are plotted in Fig. 3.

The $\gamma$ of a-Si and a-SiO$_2$ at low-frequencies is negative, and tends to 1 at high frequencies. The magnitude of their $\gamma$ is comparable, varying from -2 to 1. For a-Si$_3$N$_4$, $\gamma$ at low frequencies (<6 THz) remains constant on average, whereas at high frequencies, it fluctuates between 0.5 to 7 without showing a clear trend. The $\gamma$ of $Si_3N_4$ is several times larger than that of a-Si and a-$SiO_2$, indicating that vibration modes in a-$Si_3N_4$ are more sensitive to strains because it has stronger chemical bonds.

**IV(C). The relaxation time of propagons predicted by Akhiezer model**

Now we move on to the calculation of the average relaxation time of vibrational modes in amorphous ($\tau_{av}$) by the NMD method. The process of the NMD calculation is shown by follows:

The time-domain atomic vibration energy at the gamma point is transformed into frequency-domain spectra $\Phi(v, \omega)$ via Fourier transformation [17]:

$$\Phi(v, \omega) = \lim_{\tau \to \infty} \frac{1}{2\tau} | \frac{1}{\sqrt{2\pi}} \int_0^\tau Q(v, t) \exp(-i\omega t)\, dt|^2 \tag{12}$$

with

$$Q(v, t) = \sum_{\alpha, i} \sqrt{\frac{m_i}{N}} u_\alpha(i, t) e^*(v, i) \times \exp[i(\mathbf{0} \cdot \mathbf{r_i})] \tag{13}$$

where $m_i$ is the mass of the $i^{th}$ atom, $N$ is the total number of atoms, $u_\alpha$ is the $\alpha$ component atomic velocity, $r_i$ is the equilibrium position of the $i^{th}$ atom. $e^*$ is the conjugate of eigenvalue for vibration modes. The summation is taken over three components and all of the atoms.

The vibrational-mode frequency and linewidth ($\Gamma$) are then predicted by fitting spectral energy of each mode with the Lorentzian function as:

$$Q(v, \omega) = \frac{C_0(v)}{[\omega_0(v) - \omega]^2 + \Gamma^2(v)} \tag{14}$$

where the constant $C_0(v)$ is related to the energy of each mode.

The relaxation time of each mode $v$ is obtained by the relation as:

$$\tau(v) = \frac{1}{2\Gamma(v)} \tag{15}$$

The frequency-dependent relaxation time for a-Si, a-$SiO_2$ and a-$Si_3N_4$ are summarized in Fig. 4. To validate our calculation, we compared our results with available experimental data for a-Si and a-$SiO_2$ using inelastic x-ray scattering (IXS) [36,44], picosecond optical technique (POT) [47], Brillouin ultraviolet scattering (BUVS) [43,48], and Brillouin light scattering (BLS) [49], which shows good agreement. The magnitude of relaxation time of our results for a-Si and a-$SiO_2$ also matches that of works from Larkin *et al* [17], except that the frequency dependence for a-$SiO_2$ is different. After the validation of our calculation, the average relaxation time is evaluated as 0.913 ps for a-Si, 0.378 ps for a-SiO$_2$ and 0.164 ps for a-Si$_3$N$_4$.

With the parameters calculated in Sec. IV (Table 1 and Table 2), we are able to estimate the relaxation time of propagons via Akhiezer model (Fig. 4). The magnitude of the relaxation time ($B$) in Eq. (7) is summarized in Table 3. It is shown that our calculation not only agrees well with the NMD predictions, but also matches well with the experimental measurements for a-SiO$_2$. Moreover, The value of $B$ for a-Si agrees with that of Larkin *et al* [17], whereas for a-SiO$_2$, the value of $B$ in our calculation is 4 times larger. However, we judged that the Akhiezer model to work better as it agrees with the experimental results. We noticed that the prediction of Akhiezer starts to deviate from the results of NMD at 2 THz for a-Si, 1 THz for a-SiO$_2$ and 6 THz for a-Si$_3$N$_4$. These values are close to the transition frequency of propagons and diffusons, suggesting that the Akhiezer model is no longer valid for diffusons, whose relaxation process is governed by elastic scatterings due to local modulation of force constant from atomic disorders [50].

## V. DIFFUSIVITY AND THERMAL CONDUCTIVITY OF PROPAGONS

To further validate the calculation of the relaxation time of propagons in terms of Akhiezer model, we compare the diffusivity of propagons ($D_P$) with the results obtained from the AF theory. $D_P$ is obtained according to a comparison of Eqs. (2) and (3):

$$D_\mathrm{P} = \frac{1}{3}v_s^2 \tau(\omega) \qquad (16)$$

The results for a-Si, a-SiO$_2$ and a-Si$_3$N$_4$ are summarized in Fig. 5, which shows that the magnitude of $D_P$ obtained from Akhiezer agrees well with that obtained from AF theory. Moreover, a large part of diffusons has diffusivity that is larger than the Ioffe-Regel (I-R) limit:

$$D_\mathrm{IR} = v_s a \qquad (17)$$

where $a$ is the average bond length (Fig. 1), which is 0.19 nm for a-Si$_3$N$_4$, 0.17 nm a-SiO$_2$ and a-Si for 0.25 nm, with the corresponding D$_{IR}$ as 4×10$^{-7}$ m$^2$/s, 1.6×10$^{-7}$ m$^2$/s and 3.25×10$^{-7}$ m$^2$/s.

The total thermal conductivity of amorphous $\kappa_T$ at 300 K obtained by Eq. (1) is compared with the experimental measurements in Fig. 6. The calculated $\kappa_T$ is 1.81 W/m K for a-Si, 1.37 W/m K for a-SiO$_2$, and 2.9 W/m K for a-Si$_3$N$_4$, which agree well with the experimental measurement [51,52] and the theoretical works of Larkin *et al* [17]. We now discuss $\kappa_P$ and $\kappa_D$ shown as the blue and purple bar in Fig. 6. The value of $\kappa_P$ is 0.5 W/m K for a-Si, 0.43 W/m K for a-SiO$_2$ and 1.8 W/m K for a-Si$_3$N$_4$. The $\kappa_P$ for a-Si agrees with the result (0.63 W/m K) in the works of Larkin *et al* [17], however, $\kappa_P$ for a-SiO$_2$ predicted here is four times larger than that of Larkin [10] (0.1 W/m K). The reason is that the magnitude of relaxation time of propagons (*B*) in a-SiO$_2$ estimated by Akhiezer model is four times that of Larkin *et al* [17]. Moreover, we noticed that $\kappa_D$ of these materials is similar, which is around 1 W/m K, whereas $\kappa_P$ varies significantly depending on the materials. As a consequence, the contribution of propagons varies largely in the total thermal conductivity: $\kappa_P$ of a-Si and a-SiO$_2$ contribute to 30% of $\kappa_T$, while $\kappa_P$ of a-Si$_3$N$_4$ could contribute to as much as 70%.

## VI. CONCLUSION

In summary, we demonstrated that the relaxation process of propagons in typical amorphous materials (a-Si, a-SiO$_2$ and a-Si$_3$N$_4$) is dominated by Akhiezer mechanism. The parameters of the Akhiezer model were evaluated by using harmonic lattice dynamics and NMD methods. The appropriate sound speed of propagons is around 80% of the Debye sound speed and comparable to that of transversal sound speed. The relaxation time of propagons predicted by the Akhiezer model matches well with the numerical results from NMD and experimental measurements using IXS, BUVS, POT, BLS. Moreover, $\kappa_D$ of these amorphous are similar, which is around 1 W/m K, while the contribution of propagons varies significantly depending on the materials, which is 30% in amorphous silicon and silica but can be as high as 70% in amorphous silicon nitride.

**Acknowledgments**

This work was partially supported by CREST "Scientific Innovation for Energy Harvesting Technology" (Grant No. JPMJCR16Q5) and Grant-in-Aid for Scientific Research (A) (Grant No. 19H00744) from JSPS KAKENHI. Y.L. thanks the Fellowship (Grant No. JP18J14024) from JSPS. The calculations in this work were partially performed using supercomputer facilities of the Institute for Solid State Physics, the University of Tokyo.

**Figures**

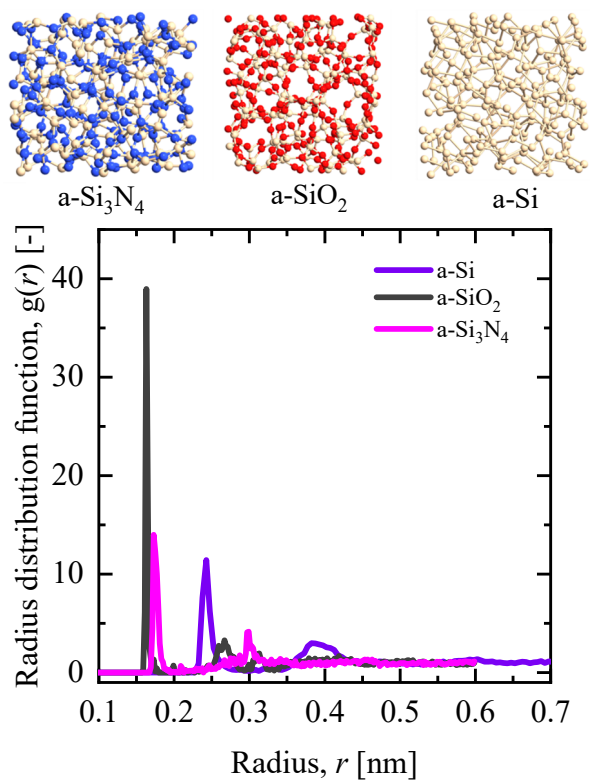

**Fig. 1** Atomistic structures and radial distribution function of a-Si, a-SiO$_2$ and a-Si$_3$N$_4$. The average bond length can be identified from the first peak of the distribution function, which ranges from 0.23 to 0.27 nm for a-Si, from 0.16 to 0.18 nm for a-SiO$_2$ and from 0.18 to 0.2 nm for a-Si$_3$N$_4$.

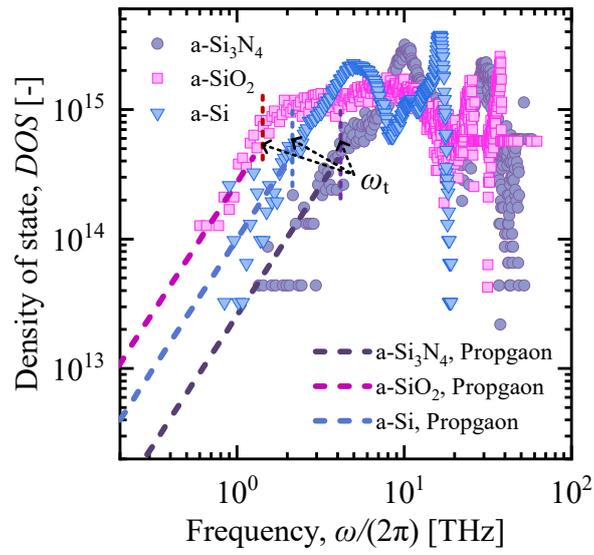

**Fig. 2** Vibrational mode density of states for a-Si$_3$N$_4$, a-SiO$_2$ and a-Si. The dashed lines are fitting curves for *DOS* of propagons with Debye approximation to determine the $v_s$.

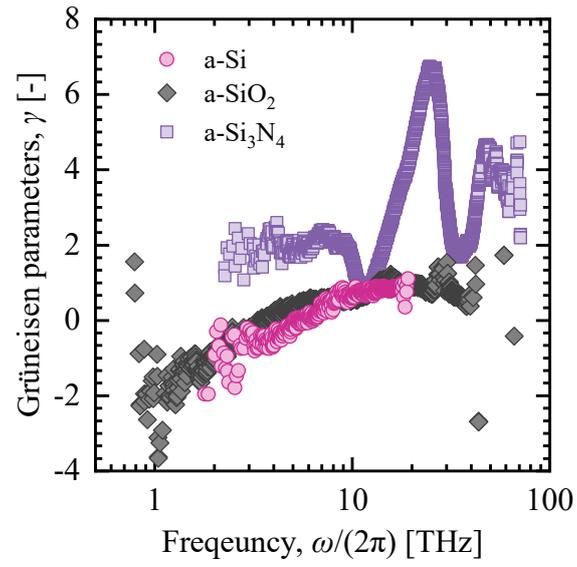

**Fig. 3** Mode-dependent Grüneisen parameters for a-$Si_3N_4$, a-$SiO_2$ and a-Si as functions of frequency.

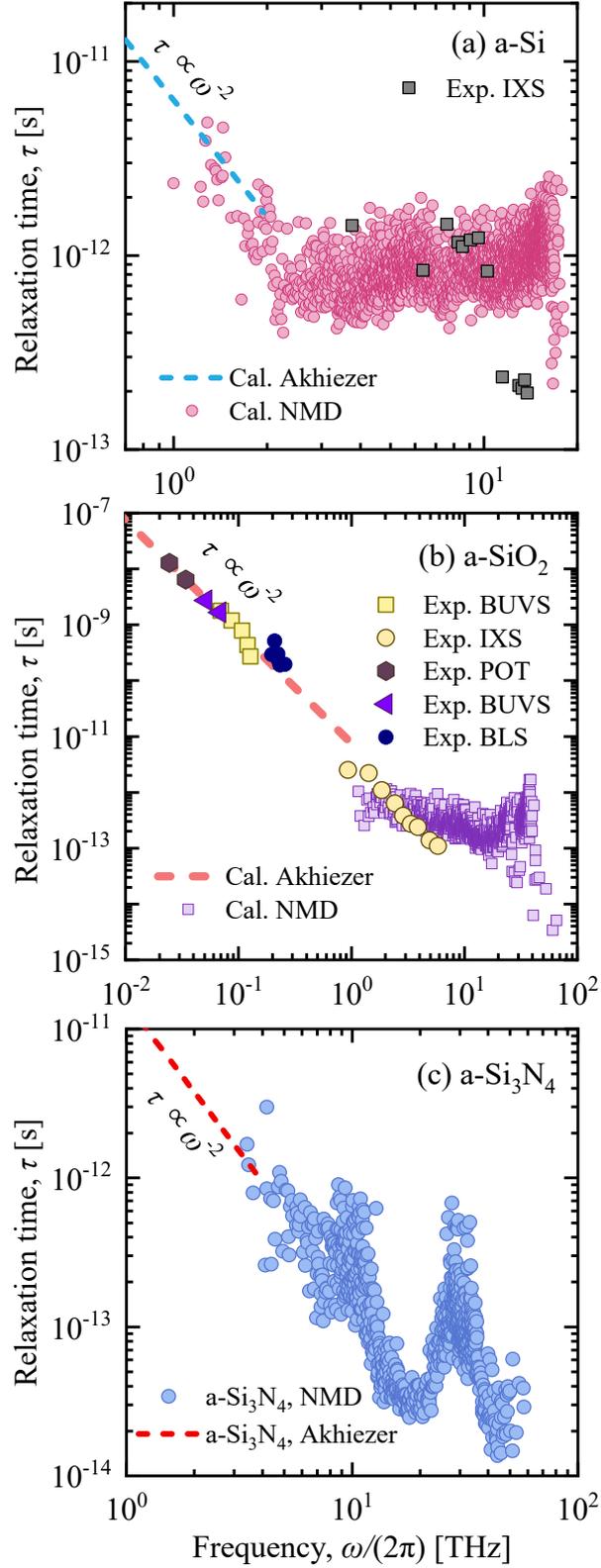

**Fig. 4** Relaxation time of amorphous materials as a function of frequency. (a) a-Si$_3$N$_4$, (b) a-SiO$_2$, (c) a-Si. The dashed lines are relaxation time predicted from Akhiezer. The experimental data in (a) and (b) for a-SiO$_2$ is obtained by IXS [36,44], POT [47], BUVS [43,48], and BLS [49].

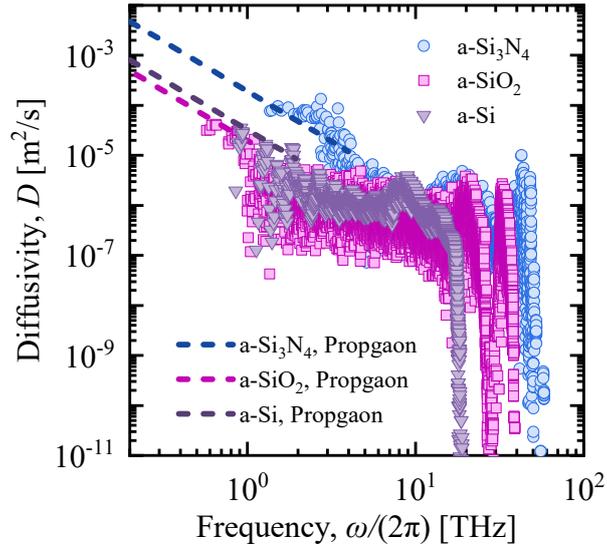

**Fig. 5** Diffusivity of amorphous materials as a function of frequency. (a) a-Si$_3$N$_4$, (b) a-SiO$_2$, (c) a-Si. The dashed lines are diffusivity of propagons.

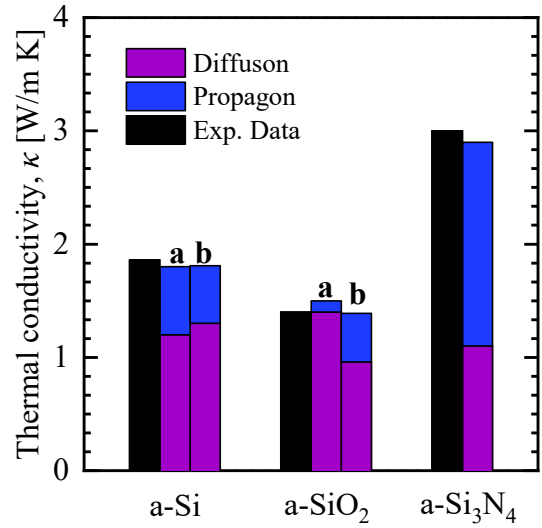

**Fig. 6** Thermal conductivity of a-Si, a-SiO$_2$, and a-Si$_3$N$_4$ at 300 K and a comparison with previous calculation works [17] and experimental data [51,52]. The symbol **a** represents the results in this work, while **b** represents the results from the works of Larkin *et al* [17].

**Tables**

Table 1 The predicted $v_s$, $S_L$, $S_T$ and $v_{dby}$ for a-Si, a-SiO$_2$ and a-Si$_3$N$_4$, and a comparison with previous theoretical works and experimental measurements.

| a-Si | $v_s$ (m/s) | $S_L$ (m/s) | $S_T$ (m/s) | $v_{dby}$ (m/s) | $(v_{bdy} - v_s)/v_{bdy}$ (%) |
|---|---|---|---|---|---|
| This work | 3915 | 7915 | 4198 | 4692 | 16.5% |
| Reference [17] | 3615 | 8047 | 3699 | 4168 | 13.3% |
| Exp. Data [36,39] | - | 7950 | 4290 | 4789 | - |
| a-SiO$_2$ | $v_s$ (m/s) | $S_L$ (m/s) | $S_T$ (m/s) | $v_{dby}$ (m/s) | $(v_{bdy} - v_s)/v_{bdy}$ (%) |
| This work | 2800 | 6200 | 3234 | 3618 | 22.6% |
| Reference [17] | 2528 | 4779 | 2732 | 3036 | 16.7% |
| Exp. Data [40-43] | - | 6060 | 3300 | 3681 | - |
| a-Si$_3$N$_4$ | $v_s$ (m/s) | $S_L$ (m/s) | $S_T$ (m/s) | $v_{dby}$ (m/s) | $(v_{bdy} - v_s)/v_{bdy}$ (%) |
| This work | 6200 | 12000 | 6680 | 7439 | 16.7% |
| Reference | - | - | - | - | - |
| Exp. Data [37] | - | 10300 | 6200 | 6857 | - |

Table 2 Summary of calculated parameters for Eq. 7. $\tau_{av}$ is calculated by using the data in Fig. 4.

| Items | $\rho$ [kg/m$^3$] | $C_v$ [kg m$^{-1}$ s$^{-2}$ K$^{-1}$] | $\tau_{av}$ [ps] | $\gamma_{av}$ [-] |
|---|---|---|---|---|
| a-Si$_3$N$_4$ | 2800 | 1.506×10$^6$ | 0.164 | 1.98 |
| a-SiO$_2$ | 2230 | 1.602×10$^6$ | 0.378 | 0.32 |
| a-Si | 2300 | 1.662×10$^6$ | 0.913 | 0.314 |

Table 3 The Calculated magnitude of relaxation time ($B$) of Eq. 7 and a comparison with Reference.

| Items | a-Si | | a-SiO$_2$ | | a-Si$_3$N$_4$ | |
|---|---|---|---|---|---|---|
| | Ref. [17] | Akhiezer | Ref. [17] | Akhiezer | Ref. | Akhiezer |
| $B$ [10$^{14}$ rads$^2$s$^{-1}$] | 2.76 | 2.5 | 0.565 | 2.97 | - | 5.96 |